\documentstyle[emulateapj]{article}

\begin{document}

\def\etal{{et al.}\ }
\def\hal{H$\alpha$}
\def\hbeta{H$\beta$}
\def\nion{$N_{ion}$}
\def\kms{km s$^{-1}$}
\def\lam{$\lambda$}
\def\per{$^{-1}$}
\def\persq{$^{-2}$}
\def\percucm{cm$^{-3}$}
\def\ncrit{$n_{crit}$}

\submitted{Accepted for publication in \emph{The Astrophysical Journal
Letters}}

\title{Polarized Broad \hal\ Emission from the LINER Nucleus of NGC
1052}

\author{Aaron J. Barth} 

\smallskip

\affil{Harvard-Smithsonian Center for
Astrophysics, 60 Garden St., Cambridge, MA 02138  \\ email:
abarth@cfa.harvard.edu} 

\bigskip

\author{Alexei V. Filippenko and Edward
C. Moran} 

\affil{Department of Astronomy, University of California,
Berkeley, CA 94720-3411  \\ email:
alex@astro.berkeley.edu, emoran@astro.berkeley.edu}

\begin{abstract}
Optical spectropolarimetry of the nucleus of the LINER NGC 1052,
obtained at the Keck Observatory, reveals a rise in polarization in
the wings of the \hal\ line profile.  The polarization vector of \hal\
is offset by 67\arcdeg\ from the parsec-scale radio axis and by
83\arcdeg\ from the kiloparsec-scale radio axis, roughly in accord
with expectations for scattering within the opening cone of an
obscuring torus.  The broad component of \hal\ has FWHM $\approx2100$
\kms\ in total flux and FWHM $\approx5000$ \kms\ in polarized light.
Scattering by electrons is the mechanism most likely responsible for
this broadening, and we find $T_e \approx 10^5$ K for the scattering
medium, similar to values observed in Seyfert 2 nuclei.  This is the
first detection of a polarized broad emission line in a LINER,
demonstrating that unified models of active galactic nuclei are
applicable to at least some LINERs.

\end{abstract}


\section{Introduction}

Low-ionization nuclear emission-line regions, or LINERs, are found in
38\% of nearby emission-line galactic nuclei (\cite{hfs97a}), yet
their physical origin remains controversial.  Observational evidence
has clearly demonstrated that some LINERs are genuine active galactic
nuclei (AGNs) powered by nonstellar processes (\cite{fil96};
\cite{ho98}), but the majority of LINERs do not show clear signatures
of AGN activity.  The interpretation of LINER spectra is confounded by
the fact that the optical emission-line intensity ratios in these
objects can often be reproduced reasonably well by models based on
AGN-like nonstellar photoionization (\cite{fn83}; \cite{hs83}), or
photoionization by hot stars (\cite{ft92}; \cite{shi92}), or fast
shocks (\cite{ds95}).  Furthermore, the weak level of the activity in
these galaxies makes it difficult or impossible to isolate any
nonstellar continuum emission that might be present in the optical
spectrum.

In a spectroscopic survey of nearly 500 nearby galaxies, Ho \etal
(1997a) found that $\sim15\%$ of LINERs have a broad component of
\hal, similar to that seen in Seyfert 1 nuclei.  By analogy with the
Seyfert population, it is convenient to define ``LINER 1'' and ``LINER
2'' subclasses based on the presence or absence, respectively, of
detected broad-line emission.  The broad-lined objects are likely to
contain genuine AGNs, but the status of the LINER 2 class is less
clear.  While some objects of this class may be AGNs in which the
central engine is either very faint or heavily obscured, in others the
narrow emission lines may be excited by processes unrelated to AGN
activity.  Since LINER 2s are potentially the most abundant class of
AGNs, it is worthwhile to determine which mechanisms are responsible
for their emission properties.

The major breakthrough in understanding the connection between the
Seyfert 1 and 2 classes was the discovery by Antonucci \& Miller
(1985) of a ``hidden'' broad-line region in the Seyfert 2 galaxy NGC
1068.  Based on such evidence, unified models of AGNs interpret the
differences between type 1 and type 2 Seyferts as the result of
orientation-dependent obscuration by an optically thick, dusty torus
surrounding the central engine.  In order to determine whether unified
models apply to LINERs, we have begun a spectropolarimetric survey of
LINERs at the Keck Observatory, and complete results of the survey
will be presented in future papers.

In this \emph{Letter}, we present our Keck observations of NGC 1052,
an elliptical galaxy at $v_r = 1470$ \kms.  NGC 1052 has long been
considered one of the prototypical LINERs, and its emission-line
properties have been the subject of numerous studies (\cite{fos78},
1981; \cite{peq84}; \cite{hfs93}).  NGC 1052 is classified by Ho,
Filippenko, \& Sargent (1997b) as a LINER 1.9; a faint broad component
of \hal\ is detected in the total flux spectrum, but only after
careful starlight subtraction and line-profile fitting
(\cite{hfsp97}).  Its X-ray spectrum is unusually flat, with an
observed 2--10 keV photon index of $\sim0.1$ and a high inferred
absorbing column of $N_H \approx 10^{23}$ cm\persq\ (\cite{ga99}).
Additional evidence for an AGN in NGC 1052 comes from VLBI radio
observations which reveal a double-sided jet emerging from the nucleus
(\cite{wro84}; \cite{jon84}; \cite{kel98}).  It is also the only
elliptical galaxy in which H$_2$0 megamaser emission has been detected
(\cite{bwh94}).

\section{Observations and Reductions}

The observations were obtained at the Keck-II telescope on 1997
December 20 UT, using the LRIS spectropolarimeter (\cite{oke95};
\cite{coh96}).  Conditions were photometric, with $\sim 1\farcs5$
seeing.  We used a 600 g mm\per\ grating blazed at 5000 \AA\ and a
slit width of 1\arcsec, yielding a spectral resolution of $\sim6$ \AA\
and a pixel scale of 1.2 \AA\ pix\per\ over the range 4300--6830 \AA.
The exposure times were 900 s with the half-wave plate at 0\arcdeg\
and at 45\arcdeg, and 450 s with the waveplate at 22\farcs5 and at
67\farcs5, for a total of 2700 s.  The spectrograph slit was oriented
east-west (P.A. = 90\arcdeg), while the parallactic angle was
35\arcdeg; this offset should have only a minor effect on the spectral
shape as the airmass was 1.2--1.3 during the observations.

The spectra were extracted with a width of 4\arcsec\ along the slit,
wavelength and flux-calibrated, corrected for continuum atmospheric
extinction and telluric absorption bands, and rebinned to 2 \AA\
pix\per.  The large extraction width was chosen in order to minimize
the noise in the continuum; in narrower extractions, noise features
appeared in the continuum polarization spectrum with amplitudes
comparable to that of the \hal\ feature discussed below.  Such noise
features in $p$ can arise from uncertainties in interpolating counts
in fractional pixels at the edges of the extraction aperture, due to
the finite sampling (0\farcs43 pixel size) of the galaxy's spatial
profile.  Polarimetric analysis was performed according to the methods
outlined by by Miller, Robinson, \& Goodrich (1988) and Cohen \etal
(1997).  Calibration of the polarization angle was done using the
polarized standard star BD +64 106 (\cite{tur90}) as a reference.
Observations of unpolarized standard stars showed a flat, well-behaved
polarization response over the observed wavelength range.  The results
for NGC 1052 are shown in Figure \ref{fig1}.  The displayed spectrum
of $p$ is the ``rotated Stokes parameter,'' calculated by rotating $q$
and $u$ to an angle in which all of the observed polarization signal
falls in a single Stokes parameter.  The ``Stokes flux,'' which shows
the polarized component of the spectrum, is the product of total flux
and the rotated Stokes parameter.

In order to measure accurate polarizations for the emission lines,
starlight was subtracted from the total flux using a spectrum of the
nucleus of NGC 3115 as an absorption-line template.  We were able to
achieve a satisfactory subtraction without the inclusion of a
featureless continuum component, and we conclude that the featureless
continuum, if present, can contribute no more than a few percent of
the total observed flux.  All emission-line and continuum measurements
were performed on the $q$ and $u$ spectra, with the results converted
to $p$ and the position angle of polarization ($\theta$) only as the
final step.

\section{Results and Discussion}

\subsection{Polarized Line Emission}

As Figure \ref{fig1} shows, NGC 1052 displays the classic signature of
a ``hidden'' broad-line region: $p$ rises above the continuum
polarization level in the broad wings of \hal, and drops again in the
core of the \hal+[\ion{N}{2}] blend where the unpolarized narrow lines
contribute most of the flux.  Although $p$ only rises by $\sim0.2\%$
above the continuum, the broad \hal\ wings have relatively high
intrinsic polarization because of their faintness in total flux.
Overall, the \hal+[\ion{N}{2}] blend has $p = 0.9\% \pm 0.1\%$, while
the \hal\ wings have $p = 4.8\% \pm 1.0\%$ over the range 6500-6530
\AA\ (rest wavelength) and $p=6.7\% \pm 1.2\%$ over 6600-6630 \AA.
The line is significantly broadened in polarized light; in Stokes
flux, the spectrum of NGC 1052 is clearly that of a broad-lined AGN.

To measure the polarization of the broad component of \hal, the
\hal+[\ion{N}{2}] blend was decomposed by fitting a set of Gaussian
components.  In total flux, fitting the blend with three narrow
Gaussians and a broad \hal\ component proved unsuccessful; the fit
presented by Ho \etal (1997c) indicates that medium-width components
of \hal\ and [\ion{N}{2}] lines are present as well.  Following the
method of Ho \etal (1997c), we fit the blend using seven Gaussians,
representing narrow and medium-width components for each line plus
broad \hal.  The widths of the narrow (FWHM = 480 \kms) and
medium-width (FWHM = 1020 \kms) components, as well as the wavelength
separation between the narrow and medium-width components for each
line, were set to match the values found by fitting the [\ion{S}{2}]
doublet.  The wavelength separation and flux ratio between the
[\ion{N}{2}] \lam6548 and \lam6583 lines were fixed to match the known
values, while the broad \hal\ width was allowed to vary freely in the
fit.  The best fit to the \hal+[\ion{N}{2}] blend closely matches that
of Ho \etal (1997c), and we find that the broad \hal\ component in
total flux has FWHM = $2120 \pm 70$ \kms\ and $f = (1.9 \pm 0.5)
\times 10^{-13}$ erg cm\persq\ s\per.  A small excess above the level
of the best-fitting decomposition appears in the \hal\ wings at
velocities out to $\sim3400$ \kms, and may be the same broadened \hal\
component that appears in the Stokes flux spectrum.

In polarized flux, the signal-to-noise ratio is too low to justify
such a detailed fit, and we simply fit the profile with four
Gaussians, representing the three narrow components and a broad \hal\
line.  The best fit indicates FWHM = $4920 \pm 450$ \kms\ and $f =
(4.5 \pm 0.4) \times 10^{-15}$ erg cm\persq\ s\per\ for broad \hal.
Figure 2 shows the \hal+[\ion{N}{2}] profiles and the fits in total
and polarized flux.  Taking the ratio of the broad-line fluxes in
polarized and total light, the polarization of broad \hal\ is $2.3\%
\pm 0.6\%$.

If electron scattering is responsible for the polarization of broad
\hal\ and the broadening of the line in polarized light, the
temperature of the scattering electrons can be estimated from the
\hal\ linewidth.  Subtraction in quadrature of the linewidth in total
flux from the linewidth in Stokes flux yields the quantity $\Delta
u_{th}$, an estimate of the thermal broadening, which can be related
to the electron temperature by $T_e \approx m_e\Delta
u_{th}^2/(16k\ln2)$ (\cite{mgm91}).  For the observed $\Delta u_{th}$
of 4440 \kms\ in NGC 1052, we find $T_e \approx 1.2 \times 10^5$ K, a
factor of $\sim2-3$ lower than the temperature found for the
scattering electrons in NGC 1068 (Miller \etal 1991).

One potential difficulty with the electron-scattering interpretation
is that a medium of scattering electrons at T $\approx 10^5$ K cools
rapidly; the cooling time for such a plasma is of order 0.8 $n_3^{-1}$
yr, where $n_3$ is the electron density in units of $10^3$ \percucm.
The power required to maintain the electrons at $10^5$ K is $\sim 3
\times 10^{40} n_3^2 V$ erg s\per, where $V$ is the volume of the
scattering region in pc$^3$.  NGC 1052 has an unabsorbed 2--10 keV
luminosity of $\sim5 \times 10^{41}$ erg s\per\ (\cite{ga99}),
sufficient to heat the scattering region provided that the scattering
medium is optically thin and confined to a small volume (within the
opening cone of a parsec-scale torus, for example).  Alternatively,
scattering by dust grains in the obscuring torus (e.g., \cite{kar95};
\cite{dop98}) rather than by electrons would alleviate any energy
budget problem, but it is not clear how dust scattering could cause
the \hal\ line to appear broader in polarized light than in total
flux.

Polarization by scattering within the opening cone of an obscuring
torus results in a polarization angle oriented perpendicular to the
rotation axis of the torus.  In NGC 1052, the best indication of the
nuclear orientation is the radio axis at milliarcsecond resolution,
which we assume is at least roughly oriented along the torus rotation
axis.  From the VLBI map of Kellermann \etal (1998), the jet axis is
oriented at P.A. $\approx67\arcdeg$, while the jet bends to a P.A. of
95\arcdeg\ at kiloparsec scales (\cite{wro84}).  The \hal+[\ion{N}{2}]
emission blend is polarized at $\theta = 178\arcdeg \pm 2\arcdeg$,
which is offset by 69\arcdeg\ from the VLBI jet axis and 83\arcdeg\
from the kpc-scale radio axis, in fair agreement with expectations for
the obscuring torus picture.

The only narrow line having $p$ significantly above that of the
continuum is [\ion{O}{1}] \lam6300, with $p = 1.0\% \pm 0.1\%$ at
$\theta = 174\arcdeg \pm 5\arcdeg$.  The other narrow lines appear in
the Stokes flux spectrum, but this is a likely result of foreground
polarization in the Galaxy.  The Galactic reddening along the line of
sight to NGC 1052 is $E(B-V)=0.027$ mag (\cite{sfd98}), implying a
maximum polarization by transmission through Galactic dust of 0.24\%
(\cite{smf75}).  Since no feature at 5007 \AA\ appears in $p$, the
most likely explanation for the appearance of [\ion{O}{3}] in Stokes
flux is a foreground screen of aligned dust grains, either in NGC 1052
or in the Galaxy, affecting both the emission lines and continuum.
Similarly, the [\ion{N}{2}] and [\ion{S}{2}] features appearing in the
Stokes flux spectrum can be attributed to foreground polarization.

In Stokes flux, the [\ion{O}{1}]/[\ion{S}{2}] flux ratio is
substantially greater than in total flux, as can be seen in Figure
\ref{fig1}, and this may be a consequence of the large difference
between the critical densities of these two transitions.  The higher
polarization of [\ion{O}{1}] could result from density stratification
in the narrow-line region (NLR), combined with obscuration of the
inner, high-density portion of the NLR.  If the obscuring torus is
large enough to surround the [\ion{O}{1}]-emitting region of the NLR,
then electron scattering within the opening angle of the torus would
cause [\ion{O}{1}] to appear polarized, while other lines emitted by
more diffuse gas at larger radii would be unpolarized.  However, the
[\ion{O}{1}] line is \emph{narrower} in polarized light than in total
flux, with FWHM = $690 \pm 130$ \kms\ and $800 \pm 20$ \kms,
respectively.  If the [\ion{O}{1}] line were polarized in the same
manner as \hal, then its profile should be similarly broadened in
polarized light. A more likely explanation for the polarization of
[\ion{O}{1}] is transmission through a region of aligned dust grains
within the inner NLR, possibly associated with the obscuring torus.
Polarization by dust transmission on NLR scales has been observed in
some Seyfert nuclei (\cite{goo92}).

\subsection{Continuum Properties}

The measured level of continuum polarization is $0.51\% \pm 0.01\%$
over the range 4400--4800 \AA, and $0.43\% \pm 0.01\%$ over 5100--6100
\AA.  The position angle of continuum polarization ($5\arcdeg \pm
1\arcdeg$ over 5100--6100 \AA) is nearly aligned with that of the
\hal+[\ion{N}{2}] blend ($178\arcdeg \pm 2\arcdeg$); the offset of
$\sim7\arcdeg$ between the line and continuum polarizations may be a
result of foreground polarization oriented at an unrelated angle.  The
intrinsic degree of polarization of the scattered continuum is unknown
because of the strong dilution by unpolarized starlight and the
uncertain contribution of interstellar polarization.  However, even if
foreground dust contributes $p=0.24\%$ to the continuum polarization,
the nonstellar continuum must have $p \gtrsim5\%$; otherwise it would
be detectable in the total flux spectrum.

A $p$ spectrum rising toward the blue is the typical signature of dust
scattering, although the detailed calculations by Kartje (1995)
demonstrate that multiple scattering by dust can lead to
wavelength-independent polarization in the optical.  However, in NGC
1052, the starlight fraction in total flux presumably drops toward the
blue end of the spectrum, and a $p$ spectrum rising to the blue would
occur even for wavelength-independent electron scattering.

Like many Seyfert 2 nuclei (e.g., \cite{wwh88}), NGC 1052 suffers from
an apparent deficit of ionizing photons relative to the number needed
to ionize the NLR.  Given the lower limit to the \hbeta\ flux of $f
\geq 1.6 \times 10^{-13}$ erg cm\persq\ s\per\ measured by Ho \etal
(1997b) in a $2\arcsec \times 4\arcsec$ aperture, and assuming an
$f_\nu \propto \nu^{-1}$ ionizing continuum, it is easy to show
(\cite{fn83}) that the nonstellar continuum, if unobscured, should
have $f_\lambda \geq 1.6\times 10^{-15}$ erg cm\persq\ s\per \AA\per\
at 4800 \AA.  This lower limit corresponds to $\sim20\%$ of the total
continuum flux at 4800 \AA, a sufficiently large fraction that it
would have been detected during the starlight subtraction procedure.
If the NLR is photoionized by the AGN, then the optical continuum is
evidently heavily obscured.  We note that the nonstellar continuum has
yet to be detected in the ultraviolet; \emph{IUE} spectra did not
reveal any nonstellar component (\cite{fos81}), and to date no
\emph{HST} ultraviolet data have been published for NGC 1052.

Rough estimates of the unobscured continuum strength can sometimes be
made for hidden type 1 AGNs (e.g., NGC 4258: \cite{wil95}; Cygnus A:
\cite{ogl97}), but these require knowledge of the covering factor and
optical depth of the scattering medium.  Lacking such information, we
are unable to determine the intrinsic continuum luminosity.
High-resolution imaging polarimetry could improve this situation by
revealing the size and morphology of the scattering region.

\section{Conclusions}

We have detected polarized continuum and broad-line emission from the
nucleus of NGC 1052, the first such detection in a LINER.  In
polarized light the spectrum of NGC 1052 is that of a ``Type 1'' AGN.
This result indicates that unified models of AGNs may be useful for
understanding the differences between type 1 LINERs and some fraction
of type 2 LINERs.  Some narrow-lined radio galaxies are known to host
obscured broad-lined AGNs (e.g., \cite{ab90}), and NGC 1052 appears to
be a local, low-luminosity example of this phenomenon.  As we continue
our survey, we hope to determine the fraction of LINERs which show
polarized broad emission lines.

\acknowledgments

The W. M. Keck Observatory is operated as a scientific partnership
among the California Institute of Technology, the University of
California, and NASA, and was made possible by the generous financial
support of the W.M. Keck Foundation.  This work was supported by NASA
grant NAG 5-3556.  Research by A. J. B. is supported by a postdoctoral
fellowship from the Harvard-Smithsonian Center for Astrophysics.

%\clearpage
 
\begin{center}
Figure Captions
\end{center}

\figcaption[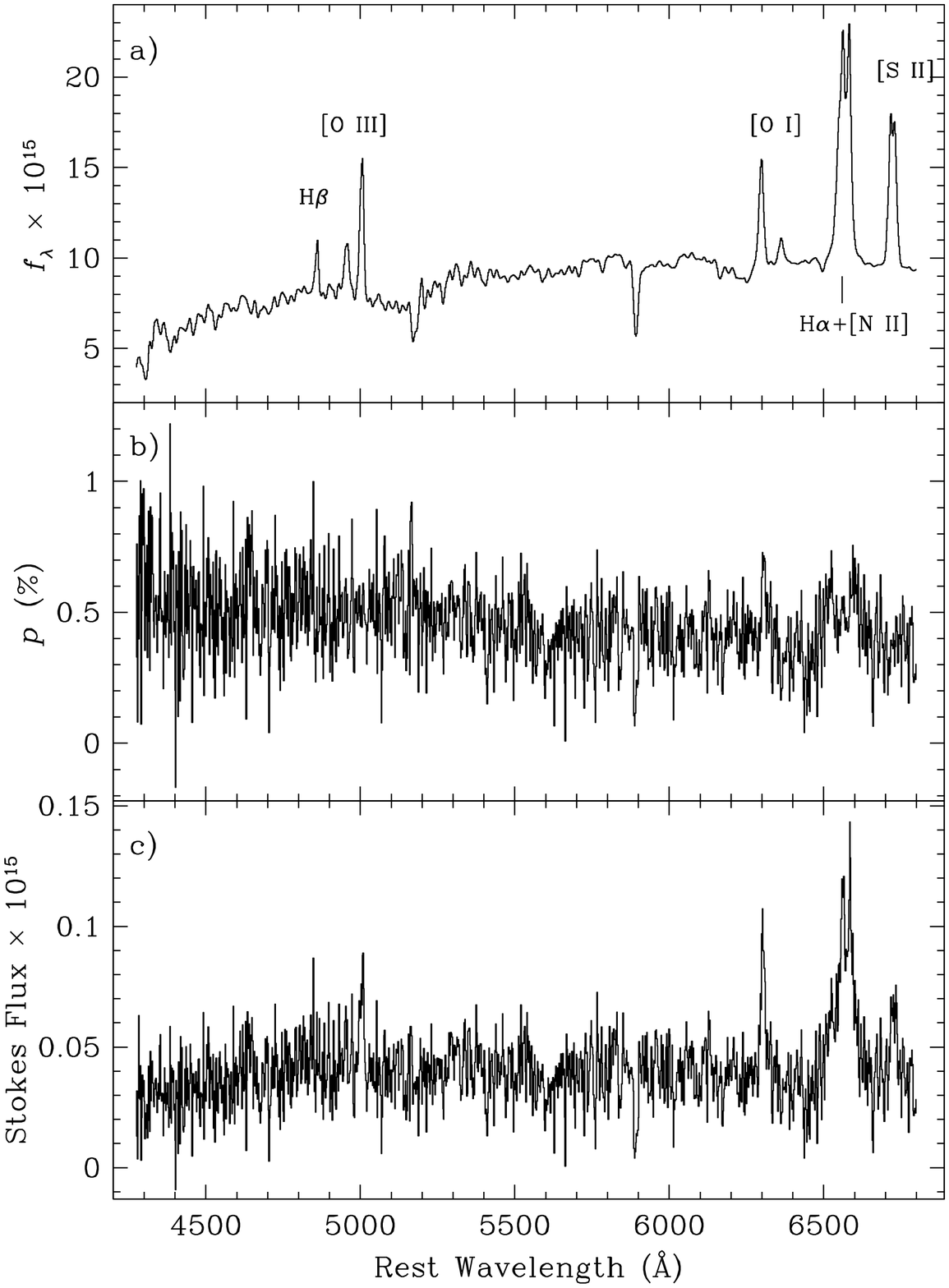]{Polarization data for NGC 1052.  {\it (a)} Total
flux, in units of $10^{-15}$ erg s\per\ cm\persq\ \AA\per.  {\it (b)}
Degree of polarization, given as the rotated Stokes parameter.  {\it
(c)} Stokes flux, equal to $p \times f_\lambda$.
\label{fig1} }

\figcaption[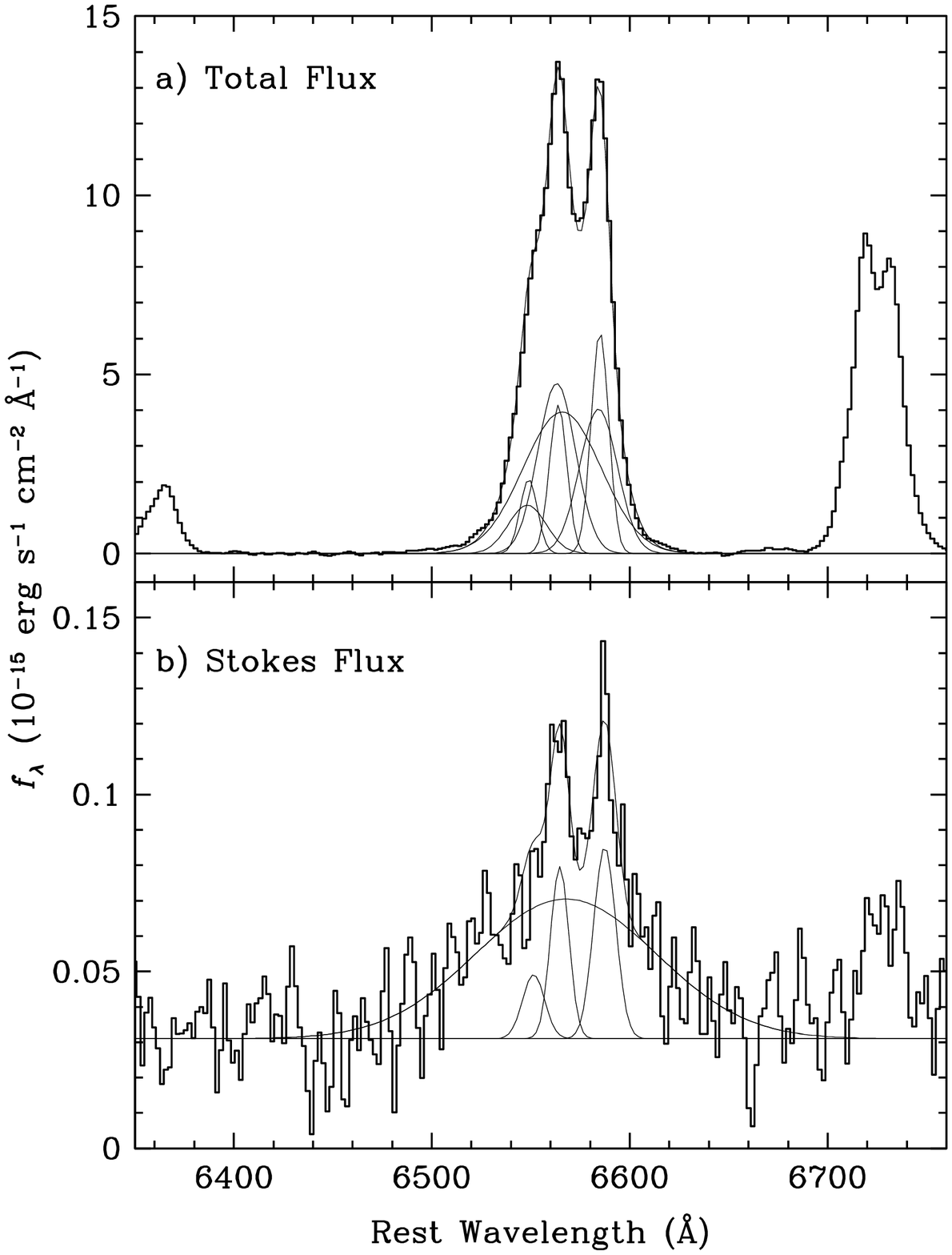]{Comparison of the \hal+[\ion{N}{2}] profiles of
NGC 1052 in total flux and Stokes flux. {\it (a)} Total flux spectrum,
after starlight subtraction, with the components of the 7-Gaussian
fit.  {\it (b)} Stokes flux spectrum and components of the 4-Gaussian
fit.
\label{fig2} }

\clearpage

\begin{figure}
\plotone{fig1.ps}
\end{figure}

\begin{figure}
\plotone{fig2.ps}
\end{figure}

\end{document}